\documentclass[iop]{emulateapj}
\usepackage{float}

\slugcomment{ver.Apr.16; draft 1 pages}
\slugcomment{ver.Aug.12; draft 6 pages}
\slugcomment{ver.Mar.26; draft 9 pages}
\slugcomment{ver.Jun.12; draft 10 pages}
\slugcomment{ver.Aug.06; draft 13 pages}
\slugcomment{ver.Aug.20; draft v1.1}
\slugcomment{ver.Sep.11; draft v2}
\slugcomment{ver.Sep.22; draft v2.1}
\slugcomment{ver.3: Oct.20; revisions after the referee's comments}
\slugcomment{Received 2014 September 22; accepted 2014 October 22}

\shorttitle{faint $z\sim6$ quasars}
\shortauthors{Kashikawa et al.}

\begin{document}


\title{The Subaru high-$z$ quasar survey: discovery of faint $z\sim6$ quasars}


\author{
Nobunari Kashikawa\altaffilmark{1,2}, 
Yoshifumi Ishizaki\altaffilmark{2}, 
Chris J. Willott\altaffilmark{3},
Masafusa Onoue\altaffilmark{2}, 
Myungshin Im\altaffilmark{4},\\
Hisanori Furusawa\altaffilmark{1}, 
Jun Toshikawa\altaffilmark{2}, 
Shogo Ishikawa\altaffilmark{2}, 
Yuu Niino\altaffilmark{1}, 
Kazuhiro Shimasaku\altaffilmark{5}, 
Masami Ouchi\altaffilmark{6},\\ 
and 
Pascale Hibon\altaffilmark{7}
}

\email{n.kashikawa@nao.ac.jp}

\altaffiltext{1}{Optical and Infrared Astronomy Division, National Astronomical Observatory, Mitaka, Tokyo 181-8588, Japan.}
\altaffiltext{2}{Department of Astronomy, School of Science, Graduate University for Advanced Studies, Mitaka, Tokyo 181-8588, Japan.}
\altaffiltext{3}{Herzberg Institute of Astrophysics, National Research Council, 5071 West Saanich Road, Victoria, BC V9E 2E7, Canada}
\altaffiltext{4}{Center for the Exploration of the Origin of the Universe (CEOU), Astronomy Program, Department of Physics and Astronomy, Seoul National University, 1 Gwanak-rho, Gwanak-gu, Seoul 151-742, Republic of Korea}
\altaffiltext{5}{Department of Astronomy, University of Tokyo, Hongo, Tokyo 113-0033, Japan.}
\altaffiltext{6}{Institute for Cosmic Ray Research, The University of Tokyo, 5-1-5 Kashiwanoha, Kashiwa, Chiba 277-8582, Japan.}
\altaffiltext{7}{Gemini Observatory, La Serena, Chile}

\begin{abstract}

We present the discovery of one or two extremely faint $z\sim6$ quasars in $6.5$ deg$^{2}$ utilizing a unique capability of the wide-field imaging of the Subaru/Suprime-Cam.
The quasar selection was made in ($i'-z_B$) and ($z_B-z_R$) colors, where $z_{B}$ and $z_{R}$ are bandpasses with central wavelengths of $8842$\AA~and $9841$\AA, respectively.
The color selection can effectively isolate quasars at $z\sim6$ from M/L/T dwarfs without the $J$-band photometry down to $z_R<24.0$, which is 3.5 mag deeper than Sloan Difital Sky Survey (SDSS).
We have selected $17$ promising quasar candidates.
The follow-up spectroscopy for seven targets identified one apparent quasar at $z=6.156$ with $M_{1450}=-23.10$.
We also identified one possible quasar at $z=6.041$ with a faint continuum of $M_{1450}=-22.58$ and a narrow Ly$\alpha$ emission with HWHM$=427$ km s$^{-1}$, which cannot be distinguished from Lyman $\alpha$ emitters.
We derive the quasar luminosity function at $z\sim6$ by combining our faint quasar sample with the bright quasar samples by SDSS and CFHQS.
Including our data points invokes a higher number density in the faintest bin of the quasar luminosity function than the previous estimate employed.
This suggests a steeper faint-end slope than lower $z$, though it is yet uncertain based on a small number of spectroscopically identified faint quasars and several quasar candidates are still remain to be diagnosed. 
The steepening of the quasar luminosity function at the faint end does increase the expected emission rate of the ionizing photon; however, it only changes by a factor of approximately two to six.
This was found to be still insufficient for the required photon budget of reionization at $z\sim6$.

\end{abstract}


\keywords{cosmology: observations --- quasars: emission lines --- quasars: general}


\section{Introduction}

High-$z$ quasars are key probes of the early universe. 
While the most distant quasars identified so far have been discovered by the UKIRT Infrared Deep Sky Survey (UKIDSS; \citealp{mor11}) and the VISTA Kilo-degree Infrared Galaxy (VIKING: \citealp{ven13}), the number of luminous quasars at $z>6$ has been significantly increased by the Sloan Digital Sky Survey (SDSS)
The high-$z$ quasars are regarded as important sites to understand the formation process of super massive black holes (SMBHs) in the early universe.
The mass of SMBHs in some of SDSS quasars at $z>6$ has been measured to be $>10^9M_\odot$.
The constraint of the quasar number density in early epochs is very important to revise formation and evolution models of early quasars/SMBHs on timescales as short as $<$1Gyr
(e.g., \citealp{li07}; \citealp{tan09}).
The first generation SMBHs are generally supposed to have formed by merging assembly of seed BHs with smaller masses \citep{vol12}.
Observational constraints on the timescale of SMBH formation may identify the seed BHs, which might be BHs as the remnants of the PopIII stars with $100M_\odot$ \citep{vol03} or BHs with $10^5M_\odot$ generated by disk instability of protogalaxies \citep{kou04}.
The quasar activity is generally thought to be maintained by mass accretion to a SMBH of $\sim10^9M_\odot$, which had to be formed within a few Gyr in the early epoch.
Several specific formation scenarios for quasars/SMBHs with co-evolving host galaxies are proposed to explain these short timescales (e.g., \citealp{vol05, lat13}); however these models will be strongly constrained by the observational measurements of SMBH mass function and Eddington ratio distribution.

Moreover, these high-$z$ quasars are among the distant bright beacons to light up many physical properties of the foreground intergalactic medium (IGM).
The SDSS studies have made an important remark that high-$z$ quasars can be used to probe the physical conditions of the IGM at high $z$ through absorption line features in the spectra, taking advantage of their intrinsic large luminosity.
The evolution of the Gunn-Peterson (GP) optical depth in SDSS high-$z$ quasars suggests that the end of the cosmic reionization process is at around $z \sim 6$ and that the cosmic neutral hydrogen fraction, $x_{HI}$, at $z > 6$ is significantly greater than zero \citep{fan06}. 
The IGM opacity was found to dramatically increase from $\tau \propto(1+z)^{4.5}$ to $(1+z)^{11}$ at $z>5.7$ by GP measurements, suggesting rapid evolution of the $x_{HI}$ in the early universe.
It is also suggested that $x_{HI}$ has a large variation from field to field, indicating a spatially patchy reionizing process (e.g., \citealp{djo06}), which might be caused by an initial large-scale structure of UV ionizing sources, such as quasars and galaxies.
However, it is still difficult to quantitatively investigate the inhomogeneity due to the extremely low number density of high-z luminous quasars.
The metal absorption lines arose in the high-$z$ quasar spectra are the only observational clue to the early history of the metal enrichment of the universe, which is closely related to the star formation history in the reionization epoch.
The IGM metallicity was found to show a possible downturn at $z>6$ \citep{rya09, sim11}, indicating rapid metal enrichment by early feedbacks from early galaxies or PopIIIs.
These observations strongly suggest that quasars at $z>6$ are extremely important to expand our knowledge of the cosmic frontier toward the so-called dark age.

As described above, some topics related to high-$z$ quasars have been advanced by SDSS, which has provided 
$\sim20$ bright quasars at $z\sim6$ over a $6600$ deg$^2$ wide-field survey \citep{fan06}.
However, the SDSS sample contains only extremely luminous active galactic nuclei (AGNs) due to the shallow limiting magnitude of the SDSS images. 
This limitation prevents us from investigating the whole shape of the AGN luminosity function, which is important to study the SMBH mass function and its evolution. 
The lack of observational evidence of the chemical evolution in the AGNs may be due to the bias toward high luminosity.
Very deep surveys targeting fainter quasars than the current observations is required to overcome these issues.

We carried out the faint quasar survey with the Subaru telescope, which has a very large field of view (FOV: $27\times34$ arcmin$^2$) camera, Suprime-Cam.
We used the $z_B$ ($\lambda_c=8842$\AA, FWHM$=689$\AA) and $z_R$ ($9841$\AA, $537$\AA) filters, which are custom-made, dividing the SDSS $z$ band into two at $9500$\AA~\citep{shi05}. 
Although the original main objective of the survey was to find quasars at $z\sim7$ with ($z_B-z_R$) vs. ($z_R-J$) color selection, $z\sim6$ quasars can simultaneously be selected out by using ($i-z_B$) vs. ($z_B-z_R$) colors.
Suprime-Cam has new red-sensitive CCDs, that have high sensitivity in the $z_B$/$z_R$ bands.
This project is a new search for the distant quasars at $z\sim6$ and $7$ that utilizes the unique capabilities of the wide-field imaging of Suprime-Cam, its high-sensitivity CCDs at $\sim1\mu$m, and $z_B$/$z_R$ filters to effectively isolate high-$z$ quasars from L/T dwarfs.
This very deep survey is complementary to other current quasar surveys, {\it e.g.}, UKIDSS/LAS \citet{ven07}, Canada-France High-$z$ Quasar Survey (CFHQS; \citealp{wil05, wil10a}), SDSS-deep \citep{jia08, jia09}, and Panoramic Survey Telescope \& Rapid Response System (PAN-STARRS; \citealp{ban14}) mainly targeting bright quasars at $z\sim6$ down to $z_{\rm lim}^{\rm AB}<22.0$.

This paper is organized as follows:
In Section 2, we describe the observation and data reduction of the survey.
In Section 3, we describe the method to select $z\sim6$ quasars.
In Section 4, we present the result of our follow-up spectroscopic observations.
After estimating the completeness of our sample as described in Section 5, we evaluate the quasar luminosity function (QLF) at $z\sim6$ in Section 6.
Several discussions of the initial SMBH formation and implications for reionization are made in Section 7, and a summary of the paper is provided in Section 8, along with a discussion of future prospects.

Throughout the paper, we assume the following cosmology parameters: $\Omega_{\rm m}=0.3$, $\Omega_\Lambda=0.7$, and $H_0=70$ $h_{70}$ km s$^{-1}$ Mpc$^{-1}$. 
These parameters are consistent with recent cosmic microwave background constraints \citep{kom11}.
Magnitudes are given in the AB system.

\section{Observation and Data Reduction}
We obtained wide and deep $z_B$/$z_R$ images with Suprime-Cam for $7$ deg$^2$ in total for three UKIDSS-Deep Extragalactic Survey (UKIDSS-DXS; \citealp{law07}) fields (^^ ^^ Lockman Hole": $10^h 57^m 00^s, +57^{\circ} 40^{\prime} 00^{\prime \prime}$ (J2000): 1.53 deg$^2$, ^^ ^^ ELAIS N1": $16^h 10^m 00^s, +54^{\circ} 00^{\prime} 00^{\prime \prime}$ (J2000): 2.55 deg$^2$, and ^^ ^^ VIMOS 4"": $22^h 17^m 00^s, +00^{\circ} 20^{\prime} 00^{\prime \prime}$ (J2000): 3.06 deg$^2$ fields). 
We retrieved the $J$ and $K$-band images from UKIDSS-DXS DR5.
Since the survey was originally designed to detect $z\sim7$ quasars as well, we selected survey fields where deep $J$-band photometry is available.   
Observations were made on the nights of UT $2009$ June 22-24.
Suprime-Cam has ten $2$k $\times 4$k MIT/LL CCDs and covers a contiguous area of $34'\times27'$ with a pixel scale of $0\arcsec.202$ pixel$^{-1}$.
Three DXS/J-band fields have been covered by $28$ FOVs of Suprime-Cam.
The transmission curves of $z_B$ and $z_R$-band filters were presented in \citet{shi05}.
The unit exposure time for each filter was determined so that the photon counts per pixel of the sky background would reach an appropriate value. 
The total integration time was $1800$ and $600$s for the $z_B$ and $z_R$ bands, respectively.
We adopted a common dithering circle pattern of a full cycle of dithering consisting of five pointings.
The sky condition was very good with a seeing size of $0.6$ arcsec.
The data were reduced using the pipeline software package SDFRED \citep{ouc04, yag02}.
The package includes bias subtraction, flat-fielding, a correction for image distortion due to the prime-focus optics, point-spread function (PSF) matching, sky subtraction, and mosaicking.
Photometric calibration was made with spectroscopic standard stars GD153 and Feige110.

The ^^ ^^ VIMOS 4" field is covered by the CFHTLS $i'$ band with the limiting magnitude of $i'^{\rm lim}=26.68$ ($3\sigma$, $2\arcsec\phi$, AB). 
The deep $i'$-band data on the ^^ ^^ Lockman Hole" and ^^ ^^ ELAIS N1" fields provided by Toru Yamada were taken by the other project, which covers almost $3.5$ deg$^2$ out of our $4$ deg$^2$ survey area.
Therefore, the total survey area to search for $z\sim6$ quasars is $6.56$ deg$^2$, limited by the available area of the $i'$-band image. 

We obtained sufficient deep images down to $z_{B}^{\rm lim}=25.55$ ($3\sigma$, $2\arcsec\phi$, AB) and $z_{R}^{\rm lim}=24.15$.
These limiting magnitudes are almost the same ($\delta$mag $=0.3$) for all the three UKIDSS-DXS fields by virtue of the stable observational condition.
We measured the limiting magnitude of UKIDSS $J$ and $K$-band images with the same procedure and under the same conditions ($3\sigma$, $2\arcsec\phi$, AB), and obtained $J^{\rm lim}=23.84$ and $K^{\rm lim}=23.17$.  
The limiting magnitude of the $i$-band image was estimated to be $i^{' \rm lim}=26.42$ and $i^{' \rm lim}=26.68$ for the Lockman/ELAIS fields and the VIMOS field, respectively.

We performed object detection and photometry by running SExtractor version 2.8.6 \citep{ber96} on the images.
Object detection was made in the $z_R$-band images.
We detected objects that had six connected pixels above $2\sigma$ of the sky background rms noise and took photometric measurements at the $2\sigma$ level.
Aperture photometry was performed with a $2\arcsec \phi$ aperture to derive the colors of the detected objects.
For all objects detected in a given bandpass, the magnitudes and several other photometric parameters were measured in the other bandpasses at exactly the same positions as in the detection-band image, using the ^^ ^^ double image mode'' of SExtractor.
Object detection and photometry were significantly less efficient and reliable close to very bright stars due to bright halos and saturation trails.
A similar degradation occurred near the edges of the images because of a low signal-to-noise ratio. 
We carefully defined ^^ ^^ masked regions" corresponding to these low-quality areas and removed all objects falling within the masked regions.
After removing the masked regions, the final effective survey area was $6.5$ deg$^2$. 

\section{Target selection}

\begin{figure}
\epsscale{1.25}
\plotone{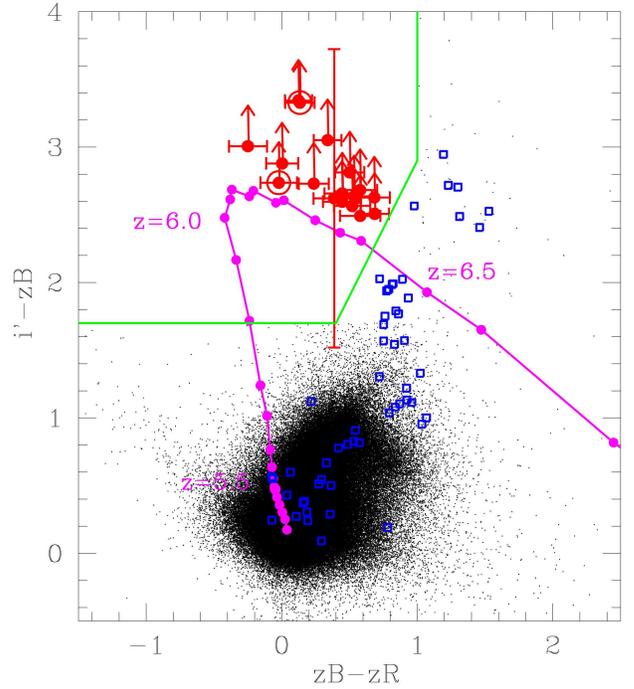}
\caption{
The ($i'-z_B$) vs. ($z_B-z_R$) color selection to identify quasars at $z\sim6$.
The quasar candidates are indicated as red circles, and lower-limit ($i'-z_B$) colors are shown when they are not detected in the $i'$ band.
Two spectroscopic identified objects are marked with large open circles.
The magenta line is the average quasar SED model with a power-law continuum and major emissions, assuming Madau's formula for continuum depression due to the IGM absorptions.
The region above the green line satisfies our color criteria of quasar candidates at $z\sim6$.
The open blue squares represent known M/L/T dwarfs from \citet{chi06}, \citet{gol04}, and \citet{kna04}. 
\label{fig_2col}}
\end{figure}

We selected quasar candidates at $z\sim6$ down to $z_R<24.15$ ($3\sigma$, $2\arcsec\phi$, AB) with $i'-z_B$ and $z_B-z_R$ color selection.
The color diagram (Figure~\ref{fig_2col}) was drawn based on one million objects detected in the $z_R$ band.
These colors were measured in a $2\arcsec\phi$ aperture.
To identify the expected position of quasars at $z\sim6$ on the color diagram and to derive the sample completeness (see Section 5), the quasar spectral energy distribution (SED) model was generated.
An alternative way to simulate the color distribution of the underlying high-$z$ quasars is to use the quasar spectra observed at low $z$ and shift them to target redshifts; however, these low-$z$ quasar samples had already passed through some specific color criteria, and we worried that a selection bias could be embedded in the sample.   
Instead, we generated model quasar SEDs, and the possible variety of quasar SEDs can be controlled by varying several model parameters.

Each quasar SED is assigned a double power-law continuum, whose average red slope at a longer wavelength than Ly$\alpha$ is $\langle\alpha_\nu\rangle=-0.79$ with a standard deviation of $\sigma_\alpha=0.34$ \citep{fan01a}, while the blue slope is $\langle\alpha_\nu\rangle=-1.57$ and $\sigma_\alpha=0.17$ \citep{tel02}.
To the continuum, we added major emission lines, whose rest equivalent width EW$_0$ is larger than 5\AA~ in \citet{van01}.
The average EW$_0$ and standard deviation of Ly$\alpha$ emission were set to $\langle$ EW$_0({\rm Ly\alpha})\rangle=69.3$\AA~and $\sigma_{{\rm EW}({\rm Ly}\alpha)}=18$\AA~\citep{fan01a}.
The FWHM and its line ratio to the Ly$\alpha$ for each emission line were kept constant following the numbers in \citet{van01}.
We did not include the Baldwin (1977) effect.
Finally, we added the continuum depression due to the intergalactic absorption blueward of Ly$\alpha$ emission following \citet{mad95}.

We have carefully selected $17$ promising quasar candidates that meet the color selection criteria of (1) $z_R<24.15$, (2) $i'-z_B>1.7$, (3) $z_B-z_R<1.0$, (4) $i'-z_B>2(z_b-z_R)+0.9$, and (5) compact stellar objects.
These color selections are reliably a long way from the dwarf star region.
The colors and image shapes of our candidates are exactly as expected for $z\sim6$ quasars.
All candidates were visually inspected to remove apparent ghosts, diffraction spikes, and spurious objects. 
Thanks to our deep image combined with many frames, we did not find contamination from cosmic rays, which was a serious problem in the SDSS quasar survey. 
Photometries in $J$ and $K$ were also taken into account, when available, for distinguishing from dwarf stars and nearby emission-line galaxies.
The $z'$-band photometry is available in the VIMOS field by CFHTLS and is also useful to discriminate from transient objects, such as supernovae, that happened to increase their luminosities during our observation epoch, though it is slightly shallower ($z'^{\rm lim}<24.6$) than our $z_B$ image.
Finally, we identified $17$ quasar candidates.
Figure~\ref{fig_2col} shows that this color selection effectively isolates quasars at $5.7<z<6.3$ from M/L/T dwarfs down to $z_{R}^{\rm lim}<24.15$, which is 3.5 mag deeper than SDSS.

\begin{deluxetable*}{llllll}
\tabletypesize{\footnotesize}
\tablecaption{Summary of Spectroscopic Observations\label{tab_obs}}
\tablewidth{0pt}
\tablehead{
\colhead{Object} & \colhead{Redshift} & \colhead{Date (UT)} & \colhead{Grism+filter} & \colhead{$T_{\rm integ}$(ks)} & \colhead{Seeing(arcsec)}  
}
\startdata
VIMOS2911001793 & 6.156 & 2012 Oct 21 & VPH850+O58 & 1200 s$\times8$ & 1.0 \\
                &       & 2013 Jun 5  & VPH900+O58 & 1200 s$\times4$ & 1.0 \\
ELAIS1091000446 & 6.041 & 2013 Mar 6\tablenotemark{a} & VPH900+O58 & 1200 s$\times3$ & 0.7 \\
                &       & 2013 Jun 5  & VPH900+O58 & 1200 s$\times12$ & 0.6 
\enddata
\tablenotetext{a}{These were identified in the course of other observational program in H. Furusawa et al. (in prep.)}
\end{deluxetable*}

\section{Spectroscopy}

Unlike other quasar surveys, our aim is to detect extremely faint quasar populations from deep imaging survey; therefore, the follow-up spectroscopy for the faint sources is challenging. 
We have not yet completed follow-up for all the candidates. 
Optical follow-up spectroscopy targeting $7$ objects out of the 17 candidates was carried out using Subaru/FOCAS \citep{kas04}.
The initial observation was scheduled in 2011; however, it has been completely canceled due to a glycol cooling accident on the Subaru telescope. 
After the recovery of the telescope, the spectroscopic observations were carried 

\begin{deluxetable*}{lllllllll}
\tabletypesize{\footnotesize}
\tablecaption{Photometry of Two Detected Objects\label{tab_obj}}
\tablewidth{0pt}
\tablehead{
\colhead{Object} & \colhead{Coordinates(J2000.0)} & \colhead{$i'$} & \colhead{$z_B$} & \colhead{$z_R$} & \colhead{$J$} & \colhead{$K$} & \colhead{$z$} & \colhead{$M_{1450}$}  
}
\startdata
VIMOS2911001793 & 22:19:17.22 +01:02:48.9 & $>27.87$       & $23.79\pm0.03$ & $23.66\pm0.11$ & $24.13\pm0.31$ & $23.15\pm0.15$ & 6.156 & $-23.10\pm 0.11$\\
ELAIS1091000446 & 16:03:49.07 +55:10:32.3 & $>27.61$ & $24.13\pm0.04$ & $24.15\pm0.13$ & $>24.69$       & $>24.36$ & 6.041 & $-22.58\pm 0.13$ 
\enddata
\end{deluxetable*}

\begin{figure}[H]
\epsscale{1.25}
\plotone{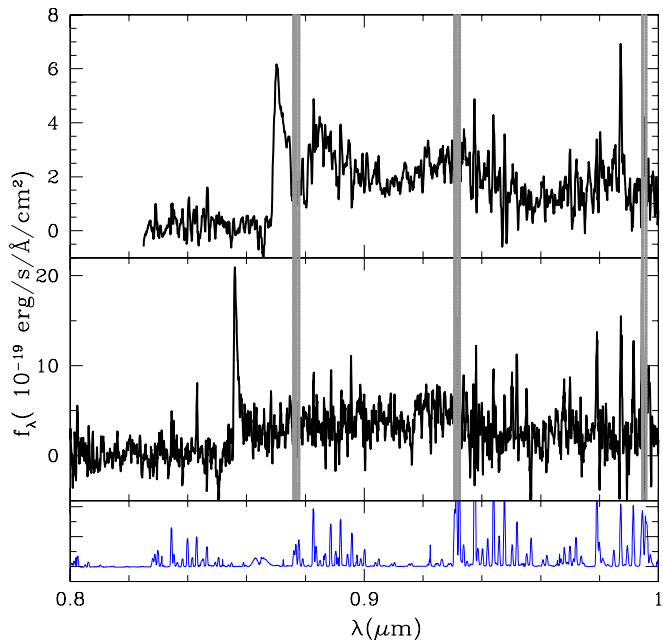}
\caption{Optical spectra of two spectroscopically identified objects at $z\sim6$. 
The upper panel shows VIMOS2911001793 and the middle panels shows ELAIS1091000446.
Hatched areas represent the wavelength ranges affected by strong night-sky lines.
The bottom figure represents sky lines.
\label{fig_spec}}
\end{figure}

\noindent out in 2012-2013, which is summarized in Table~\ref{tab_obs}.

The observation was made with either the VPH850 grating, which covers $5800$\AA$-10350$\AA~ with a pixel resolution of $1.17$\AA~, or the VPH900 grating, which covers $7500$\AA$-10450$\AA~ with a pixel resolution of $0.74$\AA.
We used the MOS mode, which allows a more secure way to align the slit quickly and accurately on such a faint target compared with the blind alignment in the long-slit mode. 
The extra MOS slits are allocated for secondary quasar candidates at $z\sim6$.
We used the long-slit mode on the 2013 March 06 run, which was carried out during the course of the other program.
We always used the $0\arcsec.8$-wide slit, which gave a spectroscopic resolution of $R\sim750$ and $R\sim1500$ for VPH850 and VPH900, respectively.

The spatial resolution was $0\arcsec.2$ pixel$^{-1}$, with two-pixel on-chip binning.
Dithering of $1\arcsec.0$ was performed during

\noindent the observation to achieve good background subtraction.
The data were reduced using standard techniques following the FOCAS data reduction pipeline.
The final spectrum was constructed from the median frame, and the flux calibration was made with spectroscopic standard star Feige 110. 

\begin{figure}
\begin{center}
\epsscale{1.15}
\plotone{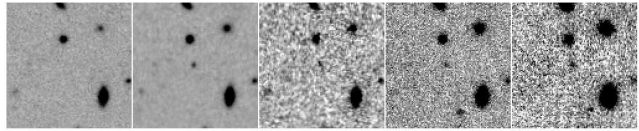}
\caption{Thumbnail images of VIMOS2911001793. 
The $i'$, $z_B$, $z_R$, $J$ and $K$-band images are shown from left to right. 
Each image is 16$\arcsec$ on a side. North is up, and east is to the left. 
\label{fig_post1}}
\end{center}
\end{figure}

\begin{figure}
\begin{center}
\epsscale{1.15}
\plotone{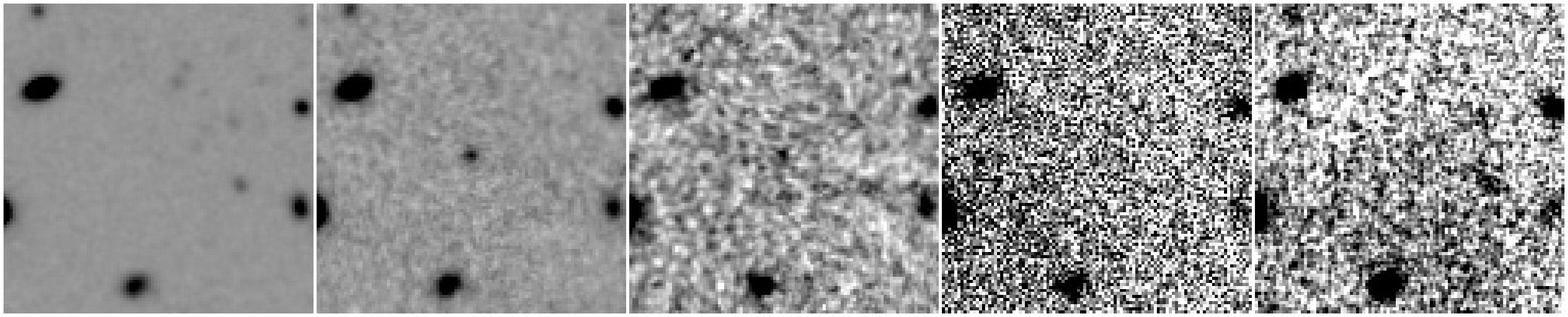}
\caption{Same as Figure~\ref{fig_post1} but for ELAIS1091000446.
\label{fig_post2}}
\end{center}
\end{figure}

Two objects, VIMOS2911001793 and ELAIS109100446, out of seven targets reveal a strong continuum break at $\sim8500$\AA.  
They have prominent asymmetric Ly$\alpha$ emissions at $8699.8$\AA~and $8559.8$\AA, corresponding to $z=6.156$ and $z=6.041$, respectively.
Additionally, both have a sharp continuum break across the Ly$\alpha$ line due to foreground IGM attenuation. 
Their $z_R$ magnitudes give $M_{1450}=-23.10$ and $-22.58$, which are much fainter than SDSS sample.
The photometries of these two objects are listed in Table~\ref{tab_obj}.
The spectra are shown in Figure~\ref{fig_spec}, and thumbnail images are shown in Figures ~\ref{fig_post1} and \ref{fig_post2}. 

\begin{deluxetable*}{llllllll}
\tabletypesize{\footnotesize}
\tablecaption{High-z Quasar Candidates\label{tab_cand}}
\tablewidth{0pt}
\tablehead{
\colhead{Object} & \colhead{Coordinates(J2000.0)} & \colhead{$i'$} & \colhead{$z_B$} & \colhead{$z_R$} & \colhead{$J$} & \colhead{$K$}  & \colhead{Notes}   
}
\startdata
Lockman14004800	& 10:51:01.24 +57:44:52.7 & $>27.61$        & $24.36\pm0.05$ & $23.68\pm0.09$ & $>25.03$  & $>24.36$ & $[$O {\sc ii}$]$ emitter \\
ELAIS891006630	& 16:09:08.08 +54:19:38.4 & $27.57\pm0.36$  & $23.86\pm0.03$ & $24.11\pm0.14$ & $>25.03$  & $24.05\pm0.41$  \\
ELAIS914002066	& 16:08:21.64 +54:57:41.2 & $>27.61$        & $24.14\pm0.03$ & $23.90\pm0.11$ & $>25.03$ & $>24.36$ \\
ELAIS914003931	& 16:08:42.46 +54:56:17.6 & $>27.61$        & $23.99\pm0.03$ & $23.99\pm0.11$ & $24.63\pm0.52$  & $>24.36$ \\
VIMOS2752003989	& 22:17:47.94 +00:06:39.6 & $>27.87$        & $24.31\pm0.04$ & $23.80\pm0.10$ & $23.72\pm0.16$  & $23.36\pm0.18$  \\
VIMOS2773005145	& 22:18:57.78 +00:27:52.8 & $27.64\pm1.31$  & $24.49\pm0.05$ & $23.81\pm0.10$ & $23.61\pm0.21$  & $23.70\pm0.25$  \\
VIMOS2833009245	& 22:21:50.01 +00:32:22.5 & $>27.87$        & $24.63\pm0.05$ & $24.05\pm0.14$ & $23.79\pm0.27$  & $23.68\pm0.26$  \\
VIMOS2853001577	& 22:22:02.28 +00:27:47.3 & $>27.87$        & $24.52\pm0.05$ & $24.08\pm0.13$ & $23.74\pm0.26$  & $23.83\pm0.30$  \\
VIMOS2733006446	& 22:18:08.83 +00:17:48.6 & $>27.87$        & $24.48\pm0.05$ & $23.95\pm0.12$ & $23.80\pm0.29$  & $>24.36$  \\
VIMOS2832005555	& 22:21:25.98 +00:04:58.0 & $27.15\pm1.06$  & $24.07\pm0.04$ & $23.73\pm0.10$ & $23.44\pm0.13$	& $23.61\pm0.17$  & ND \\
VIMOS2871008551	& 22:20:58.17 +00:06:28.8 & $27.76\pm1.86$  & $24.55\pm0.05$ & $24.03\pm0.12$ & $24.08\pm0.35$  & $>24.36$  & ND \\
VIMOS2871007103	& 22:20:46.33 +00:07:50.5 & $27.67\pm1.71$  & $24.46\pm0.05$ & $24.02\pm0.12$ & $24.20\pm0.39$  & $24.02\pm0.38$  & ND \\
VIMOS3031005637	& 22:17:18.15 +01:03:11.7 & $>27.87$        & $24.44\pm0.05$ & $23.86\pm0.12$ & $24.28\pm0.86$  & $23.74\pm0.22$  & ND \\
VIMOS2873003200	& 22:22:14.22 +00:43:06.8 & $>27.87$        & $23.78\pm0.03$ & $23.65\pm0.10$ & $23.62\pm0.22$  & $>24.36$  \\
VIMOS2993006408	& 22:18:08.34 +01:30:42.4 & $26.97\pm1.10$  & $24.35\pm0.04$ & $23.96\pm0.15$ & $23.27\pm0.23$  & $23.32\pm0.17$  
\enddata
\tablenotetext{}{The lower limit denotes the $1\sigma$ limiting magnitude.}
\end{deluxetable*}

The VIMOS2911001793 has a broad Ly$\alpha$ emission with HWHM$_{{\rm Ly}\alpha}$\footnote{This is defined here as the half-width at half-maximum of the red side of the Ly$\alpha$ emission line after correction of the spectral broadening.}=$1732$ km s$^{-1}$ and N {\sc v}$\lambda1240$ emission at $\sim8855$\AA; therefore, we conclude that this is a quasar.
It may also have a probable Si {\sc iv}+O {\sc iv} emission at $\sim9955$\AA~, corresponding to $z\sim6.142$, though the quality of the spectrum at $\sim1\mu$m is not good.  

The ELAIS109100446 has a very narrow Ly$\alpha$ emission with HWHM$_{{\rm Ly}\alpha}=427$ km s$^{-1}$, which is narrower than typical quasars except in a few cases (e.g., \citealp{gli07}, \citealp{mcg14}). 
The object has signatures of neither N {\sc v}$\lambda1240$ nor Si {\sc iv}+O {\sc iv} emissions. 
\citet{wil09} found a very faint $z\sim6$ quasar, CFHQS J0216-0455, which has $M_{1450}=-22.21$ with HWHM$_{{\rm Ly}\alpha}=800$ km s$^{-1}$, , while \citet{wil13} found the brightest $z\sim6$ Lyman break galaxies (LBGs) with $M_{1350}=-22.65$ -- $-21.33$ with HWHM$_{{\rm Ly}\alpha}=210$ -- $240$ km s$^{-1}$.
The absolute magnitude and HWHM$_{{\rm Ly}\alpha}$ of this object are between the two. 
The number density of quasars and LBGs at $z\sim6$ is almost the same at this magnitude.
It has an apparent Ly$\alpha$ emission at $z=6.041$, though any ISM absorptions lines, which are often seen on the UV spectrum of LBGs, cannot be identified.
The object is not resolved in $z_B$ and $z_R$ images under the $0\arcsec.6$ seeing size; a better PSF image is required to exactly distinguish quasars from galaxies. 
It is very likely that it is a quasar, but we cannot completely rule out the possibility that this object is a Ly$\alpha$ emitter(LAE), and the following discussions are made in two cases; our spectroscopically confirmed quasar sample is either one or two.  
Future NIR spectroscopy to detect C {\sc iv}, C {\sc iii}, and Mg {\sc ii} lines from the object will be required to make a definitive conclusion. 

The photometries of the other $15$ candidates are summarized in Table~\ref{tab_cand}.
One object, Lockman14004800, was found to be a nearby galaxy, whose  $[$O {\sc ii}$]$  emission at $z=1.2$ was identified by the spectroscopy.
We did not detect any signals from the other four spectroscopic targets (denoted as ND in Table~\ref{tab_cand}).
These may be quasars with very faint Ly$\alpha$ emission and continua or contaminants of M/L/T-dwarf stars.
We discuss the possibilities in Section 7.1.

\section{Completeness}

We used the quasar SED model described in Section 3 to estimate the sample completeness as a function of $M_{1450}$ and $z$ through a Monte Carlo simulation.
We assumed that the SEDs of faint quasars were not significantly different from those of bright ones, having a Gaussian random distribution of $\alpha_\nu$ and EW$_0$ with mean value and 1$\sigma$ uncertainties as described in Section 3.
Then, a series of artificial quasars were generated over $-27<M_{1450}<-21$ and $4.0<z<7.4$ with a step of $\Delta M_{1450}=0.5$ and $\Delta z=0.1$ for every SED model. 
The predicted ($i'$, $z_B$, $z_R$) magnitudes can be calculated from each object by converting these transmission curves.
Artificial quasars are assumed to have the same PSF size as the observation for each band and are randomly distributed on the original images after adding Poisson noise according to their magnitudes.
We used the image of the ELAIS N1 field here to estimate the completeness.
The image qualities of the three UKIDSS-DXS fields are so uniform that the same results can be expected to be derived in the other two fields.
We neglected an artificial quasar if it was within the masked regions or was heavily blended with real objects so as to evaluate the detection efficiency on sky background regions.
The object detection, photometry, and color selection were done as we did for real objects, and the fraction of objects that passed through the procedure to the input sources was estimated for a given luminosity and redshift.
We generated $800$ thousand artificial quasars in total to obtain statistically robust results.

\begin{figure}
\epsscale{1.25}
\plotone{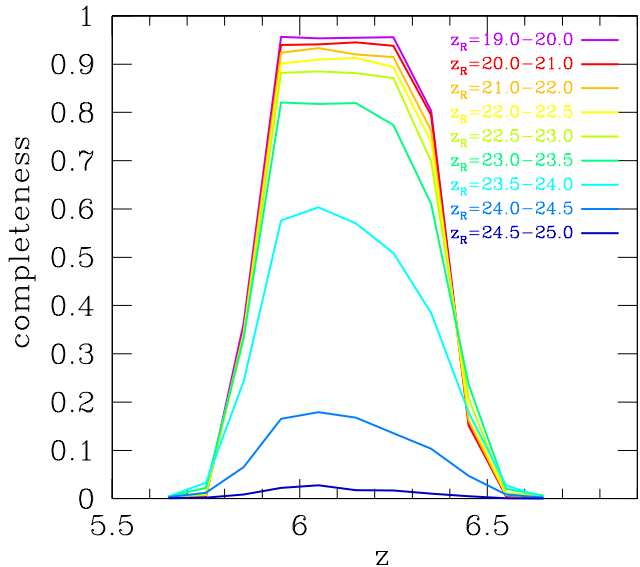}
\caption{Sample completeness as a function of redshift for each apparent $z_R$ magnitude.
\label{fig_comp_z}}
\end{figure}

\begin{figure}
\epsscale{1.25}
\plotone{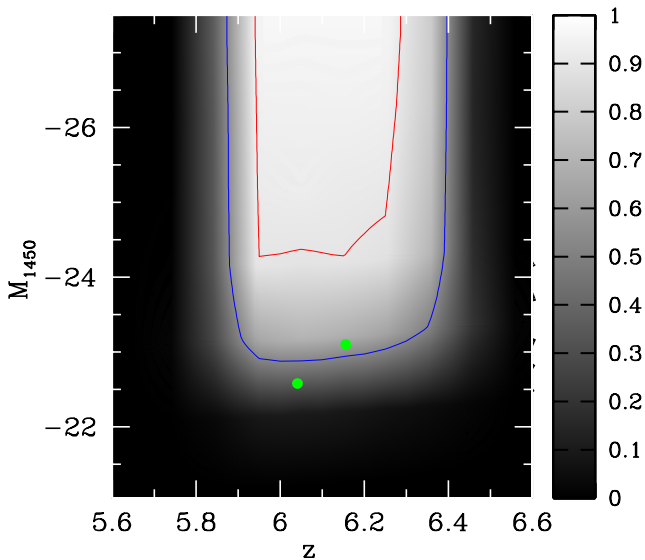}
\caption{Completeness as a function of absolute magnitude $M_{1450}$ and redshift for our sample. 
The greyscale represents the completeness from 0.0(black) to 1.0(white).
The red and blue contours show the 90\% and 50\% completeness, respectively.
The locations of two spectroscopic identified objects (left: ELAIS1091000446; right: VIMOS2911001793) in this study are also plotted.
\label{fig_comp_mz}}
\end{figure}

Figure~\ref{fig_comp_z} presents the resulting completeness estimates as a function of the apparent magnitude.
Our sample was found to be over $60\%$ complete down to $z_R=24.0$, which is close to one of our quasar selection criteria.
Figure~\ref{fig_comp_mz} presents the selection function, $p(M_{1450}, z)$, which is used to derive the effective volume when calculating the QLF.

\section{Constraint on the quasar luminosity function}

The current estimate of QLF at $z\sim6$ (Figure 3 of \citealp{wil10a}) revealed a possible break at $M_{1450}\sim-25$; however, the faintest data points at $M_{1450}\sim-22$ came from a single quasar found in the CFHQS Deep/SXDS field.
Although our spectroscopically identified sample is only one or two, the sample gives a unique independent verification of the shape of the QLF at $z\sim6$.
Combined with the faintest CFHQS quasar, our faint quasar in the $M_{1450}\sim-23.0$ bin will make a stronger constraint on the QLF faint-end slope.
Uncertainties of the quasar number density due to possible candidates without follow-up spectroscopy will be discussed on Section 7.1.

We derive the QLF at $z\sim6$ quasars by combining our faint quasar sample with the bright quasar sample provided by SDSS and CFHQS.
First, we derive the number density of our faint $z\sim6$ quasars at $M_{1450}=-22.84\pm0.5$ using the $1/V_a$ binned method \citep{ab80}.
The effective volume for a source $j$ in a $\Delta M_{1450}$ bin size, corrected with sample completeness, $p(M_{1450}, z)$, which has been derived in Section 5, can be derived from

\begin{equation}
V_a^j=\int\!\!\!\int p(M_{1450}, z)\frac{dV}{dz}dzdM_{1450},
\end{equation}

where $dV/dz$ is the comoving volume. We use only one ($M_{1450}$, $z$) bin for our small sample.
The number density $\Phi(M_{1450})$ is calculated from

\begin{equation}
\Phi(M_{1450})=\sum^N_{j=1}\frac{1}{V_a^j}(\Delta M_{1450})^{-1}.
\end{equation}

\begin{figure}
\epsscale{1.35}
\plotone{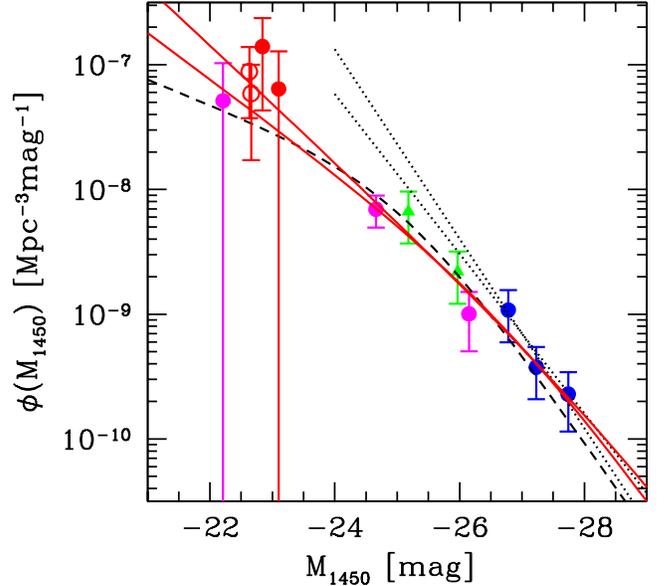}
\caption{Quasar luminosity function at $z\sim6$. 
The red filled circles at the faintest bin show the estimate of this study for case 1(right) and case 2(left) detections, while the red open circles at the faintest bin show the estimate for case 3(right) and case 4(left) (see the text for details). 
The red solid lines show the best-fit QLFs in case 1(lower) and case 4(upper), respectively.  
Blue circles, green triangles, and magenta circles are data from SDSS, SDSS-deep\citep{jia09}, and CFHQS\citep{wil10a}, respectively.
The SDSS data have been rebinned by \citet{wil10a} and converted to the cosmology in this study.
The dashed line is the double power-law fit of \citet{wil10a}, and the two dotted lines are the power-law fits of \citet{jia09}. 
\label{fig_qlf}}
\end{figure}

Figure~\ref{fig_qlf} shows the QLF compared with the SDSS and CFHQS.
The number density at $M_{1450}=-23.10\pm0.5$ is $(6.4\pm6.4)\times10^{-8}$ Mpc$^{-3}$ mag$^{-1}$ when assuming that only VIMOS2911001793 is a quasar, whereas it is $(1.4\pm0.97)\times10^{-7}$ Mpc$^{-3}$ mag$^{-1}$ at $M_{1450}=-22.84$ when assuming that both VIMOS2911001793 and ELAIS109100446 are quasars.
Our data points, even assuming that Elais109100446 is a real quasar, is consistent within $1\sigma$ errors of the CFHQS estimate.
As suggested by \citet{wil10a}, there is an evidence for a flattening of the luminosity function toward low luminosity, and our estimate is far below the extrapolation of the single power law of \citet{jia09}.

\begin{figure}
\epsscale{1.3}
\plotone{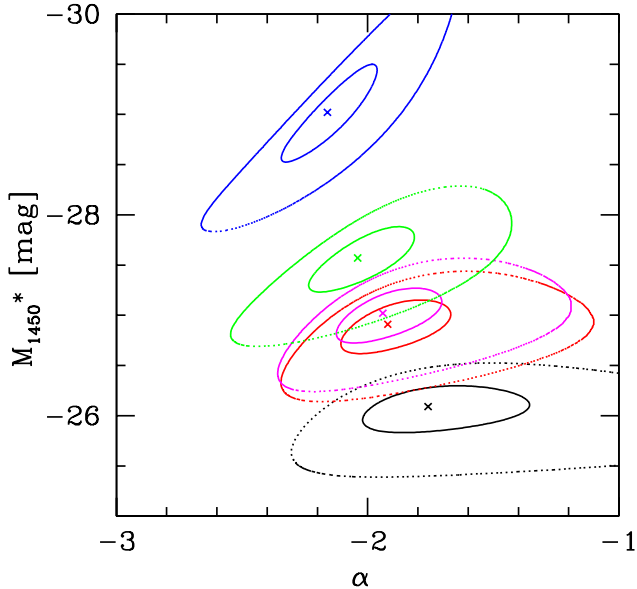}
\caption{Error ellipses of the best-fit parameters $\alpha$ and $M_{1450}^*$ for case 1(red), case 2(magenta), case 3(green), and case 4(blue; see the text for details).
The black contour shows the error ellipse in the case of only the SDSS+CFHQS sample without our data points.
The inner and outer solid ellipses are the 1$\sigma$ and 3$\sigma$ confidence levels, respectively. 
\label{fig_qlf_err}}
\end{figure}

The QLF, $\Phi(M_{1450}, z)$, is usually approximated by a double power-law:

\begin{equation}
\Phi(M_{1450}, z)=\frac{10^{k(z-6)}\Phi(M_{1450}^*)}{10^{0.4(\alpha+1)(M_{1450}-M_{1450}^*)}+10^{0.4(\beta+1)(M_{1450}-M_{1450}^*)}},
\end{equation}  

where $\alpha$ is the faint-end slope, $\beta$ is the bright-end slope, and $M_{1450}^*$ is the knee luminosity at which the slope of the QLF changes.
We follow other studies (e.g., \citealp{wil10a}; \citealp{mcg13}) in adopting $k=-0.47$, which is the evolution parameter derived from the bright end of the QLF from $z=3$ to $z=6$ \citep{fan01b}.
We also fix the bright-end slope, which can be poorly constrained by our sample, to be $\beta=-2.81$ \citep{wil10a}.
Given the very limited luminosity range of our sample, we simply fit the double power-law function to the QLF measurements at $z\sim6$ from this study, CFHQS, and SDSS surveys shown in Figure~\ref{fig_qlf}, to constrain the two parameters of $\alpha$ and $M_{1450}^*$.
We summarize the best-fit ($\alpha$, $M_{1450}^*$) parameters in Table~\ref{tab_qlf} for four cases: (1) SDSS+CFHQS+this study assuming only one quasar detection, (2) SDSS+CFHQS+this study assuming two quasar detections, (3) SDSS+CFHQS ($M_{1450}<-24$) with the faintest bin replaced with two detections in CFHQS+this study, and (4) SDSS+CFHQS ($M_{1450}<-24$) with the faintest bin replaced with three detections in CFHQS+this study. 
Figure~\ref{fig_qlf_err} shows the error ellipses of these parameters for fixed $\beta=-2.81$ at the $1\sigma$ and $2\sigma$ confidence levels.
The best-fit $M_{1450}^*$ of SDSS+CFHQS with fixed $\beta=-2.81$ in this study is significantly different from CFHQS only \citep{wil10a} of $M_{1450}^*=-25.13$, which assumes fixed $\alpha=-1.5$.
They obtained the best-fit parameters as $\beta=-3.26$ and $M_{1450}^*=-26.39$, which is close to those in this study, when assuming a steeper faint-end slope of $\alpha=-1.8$.
Including our data points invokes a higher number density in the faintest bin than the previous estimate, making steeper $\alpha$ and brighter $M_{1450}^*$.
These two parameters are correlated with each other. 
$M_{1450}^*$ tends to become brighter as $\alpha$ becomes steeper when fixing the bright-end slope \citep{mcg13}.
Our results from all four cases suggest a steep faint-end slope $\alpha\sim-2$, which is steeper than the local value $\alpha=-1.5$.
Whereas, the knee luminosity $M_{1450}^*$, which is in the range of the bright magnitude $<-25$, can be strongly constrained by precise measurements of the bright part of the QLF at $z\sim6$ by further observations.  
In case (4), $M_{1450}^*$ becomes brighter than $-28$, and $\alpha$ becomes very steep close to $\beta=-2.8$, suggesting that the QLF is no longer approximated by a double power law and instead could be almost represented by a single power law.  
It should be noted that the QLF at $z\sim6$ is still not strongly constrained because wide differences in $\alpha$ and $M_{1450}^*$, as can be seen in Figure~\ref{fig_qlf_err}, were derived from cases 1-4, whose sample difference is very small.
The faint-end slope of QLF at $z<3$ has been found to be $\alpha=-1.5$ \citep{cro09, sia08, ric05, hun04}, while a steeper value $\alpha=-1.7$--$-1.8$ is more likely at a higher redshift \citep{mcg13, ike12, gli10, mas12}, though uncertainties are large.
Our study is consistent with a steep faint-end slope of $\alpha\sim-2$.

\begin{deluxetable}{cccc}
\tabletypesize{\footnotesize}
\tablecaption{Best-fit QLF Parameters for Quasars at $z\sim6$\label{tab_qlf}}
\tablewidth{0pt}
\tablehead{
\colhead{case} & \colhead{$\phi^*$ ($10^{-9}$Mpc$^{-3}$mag$^{-1}$)} & \colhead{$M^*_{1450}$} & \colhead{$\alpha$}   
}
\startdata
1  &  $1.21^{+0.40}_{-0.37}$ & $-26.91^{+0.28}_{-0.24}$ & $-1.92^{+0.24}_{-0.19}$ \\
2  &  $1.05^{+0.36}_{-0.33}$ & $-27.02^{+0.29}_{-0.25}$ & $-1.94^{+0.22}_{-0.18}$ \\
3  &  $0.51^{+0.22}_{-0.18}$ & $-27.57^{+0.33}_{-0.31}$ & $-2.04^{+0.22}_{-0.19}$ \\
4  &  $0.08^{+0.05}_{-0.03}$ & $-29.02^{+0.47}_{-0.50}$ & $-2.16^{+0.19}_{-0.18}$   
\enddata
\end{deluxetable}

\section{Discussion}

\subsection{The Quasar Candidates}

It should be noted that we still have $10$ quasar candidates without follow-up spectroscopy.
In that sense, the current estimate of the QLF faintest bin might be the lower limit. 
Further observation of these targets is required to put stronger constraints on the faint-end slope.   
The candidate list could contain either real quasars or an intermediate population between quasar and galaxy, such as ELAIS109100446.
Otherwise, they could be contaminants of M/L/T-dwarf stars, which are either the faintest at the low-mass end of the stellar mass function or the most distant brown dwarfs detected by our deep and wide survey. 
If they were only detected in the $z_B$ and $z_R$ bands, such as ELAIS914002066, they may be transient objects. 

The follow-up spectroscopy was preferentially carried out for the most likely candidates, evaluated based on brightness in $z_B$ and NIR colors, and two-thirds of them were found not to be apparent quasars, suggesting that most of the rest were also not. 
Figure~\ref{fig_2col} shows that two spectroscopically identified $z\sim6$ objects have bluer color in ($z_B-z_R$) than most of the others.
The large clump of candidates with ($z_B-z_R$) $\sim0.5$ to $0.7$, in which three out of four ND objects are included, were close to the stellar locus and could be scattered dwarf stars. 
Lockman14004800, which was found to be an apparent $[$O {\sc ii}$]$ emitter, also has a relatively red ($z_B-z_R$) color of $0.68$.
As for the ND objects, if they were quasars, we could have instantly identified the Ly$\alpha$ emission and possibly a strong break in continuum, whereas brown dwarfs would have a faint continuum across the full wavelength range.
Actually, apparent Ly$\alpha$ emission was suddenly identified even for the first 20 minute exposure on our two spectroscopically confirmed $z\sim6$ objects.
We usually moved to the next target once a target was found to have neither an apparent continuum break nor emission after 1.5 hr integration.
Such an observational strategy would make it difficult to identify a possible faint continua from dwarf stars or line-less quasars \citep{she10, ban14}.

Many candidates of faint $z\sim6$ quasars are still waiting for spectroscopy.
Although the faintness of our sample prevents us from vigorously executing follow-up spectroscopy, more strong constraints on the faint-end QLF at $z\sim6$ would be important to constrain the BH evolution models.

\subsection{The BH Mass of Faint High-$z$ Quasars}

It is quite important to observationally determine the faint-end slope of QLF at high $z$, as several BH evolution models make qualitatively different predictions (e.g., \citealp{sha09, hop07}).
Faint quasars, which trace SMBHs with a lower mass of $<10^8 M_\odot$, are presumed to be in an early stage of SMBH growth.
Almost all bright quasars at $z=6$ are found to be accreting at approximately the Eddington limit \citep{jia07, kur09, wil10b}; however, the current BH mass measurements at $z=6$ are limited by $M_{1450}<-24.28$
Assuming the Eddington accretion ($\lambda=1$), our faint quasar, VIMOS2911001793, with $M_{1450}=-23.10$ corresponding to $L_{3000}\sim10^{38}$W, would have a BH mass of $M_{\rm BH}=5\times10^7 M_\odot$.
This is close to the BH mass, at which the highest $\lambda$ is predicted \citep{deg12}.
It is determined by the balance between the gas density available for fueling BH growth and the effective shock heating feedback from AGNs in the early universe.
Direct detection of Mg {\sc ii} emission from faint quasars is required to derive an accurate estimate of their BH masses and Eddington ratios \citep{vo09, she11}.  
It would be difficult to detect the Mg {\sc ii} emission, whose FWHM might be as narrow as $\sim1000$km s$^{-1}$ according to the tight relation between FWHM and $L_{3000}$ of $z\sim6$ quasars, from the object; however, it is interesting to assess whether the systematic high-$\lambda=1$ trend found in luminous quasars is also valid for low-luminous quasars to directly address the growth of BHs at early times.  
\citet{sha10} predict the QLF over $3<z<6$ from the growth of the BH mass function combined with the halo mass and duty circle inferred from quasar clustering measurements \citep{she07}.
This QLF model is consistent with observations even at $z=6$; however, it overpredicts the faint end of the QLF below $L\sim10^{46}$ erg s$^{-1}$, which corresponds to $M_{1450}\sim-25.5$.
This may be due to the assumption of a constant Eddington ratio and radiative efficiency over a wide luminosity range \citep{sha10}.
The higher amplitude of the faint-end QLF inferred from this study would relax the discrepancy, and future NIR spectroscopy to directly measure the $\lambda$ for the faint quasars will help to revise the model.

\subsection{Reionization}

The derived QLF at $z\sim6$ is also interesting in terms of the estimate of the quasar contribution to the photon budget of cosmic reionization.
Although quasars are expected to be the main contributor at the bright end of the luminosity function of ionizing sources, the quasar population alone cannot account for the entire requirement of ionizing photons \citep{mei05, bh07, wil10a}.
Our finding in this study of possible steepening of the QLF faint-end slope implies a greater contribution to the photon budget than in the previous estimates.

The emission rate of the hydrogen ionizing photon, $\dot{N}_{\rm ion}$ s$^{-1}$ Mpc$^{-3}$, was evaluated based on \citet{bh07}. 
Here, we consider the comoving quasar emissivity at the Lyman limit, $\epsilon^q_L=\int_{L_{\rm min}}^{\infty} L_\nu\phi(L_\nu, z)dL_\nu$ erg s$^{-1}$ Hz$^{-1}$ Mpc$^{-3}$, which is directly related to the change in the QLF, and follow the same assumptions with BH07 on the spectral index at $\lambda<912$\AA~($-1.5$), the escape fraction of ionizing photon ($f_{\rm esc}\sim1$), the quasar spectral energy distribution, and the frequency dependence of the photoionization cross-section.  
The ionizing photon density was found to be $\dot{N}_{\rm ion}=1.04\times10^{49}$ s$^{-1}$ Mpc$^{-3}$ and $1.29\times10^{49}$ s$^{-1}$ Mpc$^{-3}$ in our QLF estimate for case(1) and case(4), respectively.
The estimate is a factor of approximately two higher than the \citet{wil10a}, which is attributed to a steeper faint-end slope. 
The emissivity also depends on the minimum quasar luminosity, $L_{\rm min}$. 
We assume $M_{\rm min}(1450)=-22$ in the above calculation.
Although it is not clear how a faint AGN population would exist, if we integrate the QLF to $M_{\rm min}=-18$, $\dot{N}_{\rm ion}$ changes to $1.43\times10^{49}$ s$^{-1}$ Mpc$^{-3}$ and $2.78\times10^{49}$ s$^{-1}$ Mpc$^{-3}$ for cases (1) and (4), respectively. 
The steepening of the QLF faint-end slope increases the expected contribution of quasars to the photon budget of reionization; however, it only changes by a factor of approximately two to six based on our current QLF estimate. 
The estimate was found to be still insufficient for the required photon rate density to keep the balance with the hydrogen recombination of $\dot{N}_{\rm ion}\sim10^{50.3}$ s$^{-1}$ Mpc$^{-3}$, assuming the IGM clumping factor $C=3$.
The quasar contribution to the required photon budget is $5\%-15\%$. 
On the other hand, star-forming galaxies, which are more dominant than quasars at the limiting magnitude of $M_{\rm min}=-18$, such as LBGs \citep{fin12} and LAEs \citep{kas11}, could sustain the universe fully ionized at $z=6$ if the escape fraction of ionizing photons is larger than 0.3.

In the earlier universe, BHs in the center of each galaxy are not well matured. 
Such relatively less massive BHs will produce intense and hard FUV/EUV emission compared with usual quasars; thus, these mini-quasars may have contributed UV background radiation \citep{kaw03, mad04}.
The slope in the EUV band plays an important role in indicating the extent to which EUV photons penetrate into
primordial gas clouds against self-shielding (e.g., \citealp{tu98}).
However, the constraints from the soft X-ray background could decline this possibility (\citealp{dij04}, but see the discussion in \citealp{mei05}). 
This deep observation to probe to the faint end of the QLF will make a stronger constraint on the photon budget at $z\sim6$.

\section{Conclusion}

We have carried out an extremely faint $z\sim6$ quasar survey with Subaru/Suprime-Cam for $6.5$ deg$^2$ in total for three UKIDSS-DXS fields.
Our results are summarized as follows.

1. The quasar selection was made in ($i'-z_B$) and ($z_B-z_R$) colors, where $z_{B}$ and $z_{R}$ are bandpasses with central wavelengths of $8842$\AA~and $9841$\AA, respectively.
The color selection can effectively isolate quasars at $z\sim6$ from M/L/T dwarfs even without the $J$ band down to $z_R<24.0$, which is 3.5 mag deeper than SDSS.
We selected $17$ promising quasar candidates from the survey area, and the follow-up spectroscopy identified two objects that had a strong continuum break at $\sim8500$\AA~and prominent asymmetric Ly$\alpha$ emission.  

2. The VIMOS2911001793 has an apparent quasar spectral signature of broad Ly$\alpha$ emission with HWHM$_{{\rm Ly}\alpha}$=$1732$ km s$^{-1}$ and N {\sc v}$\lambda1240$ emission at $\sim8855$\AA.
It is at $z=6.041$, and its $z_R$ magnitude gives $M_{1450}=-23.10$.

3. The ELAIS109100446 has a very narrow Ly$\alpha$ emission with HWHM$_{{\rm Ly}\alpha}$=$427$ km s$^{-1}$.
The object has signatures of neither N {\sc v}$\lambda1240$ nor Si {\sc iv}+O {\sc iv} emissions. 
The absolute magnitude ($M_{1450}=-22.58$) and HWHM$_{{\rm Ly}\alpha}$ of this object are marginal between faint quasars and bright galaxies.
The object is not resolved under the $0\arcsec.6$ seeing size.
The object is likely to be a quasar; however, we cannot obtain any conclusive evidence, and the following discussion examines two cases; our spectroscopic confirmed quasar sample is either one or two.  

4. We derived the QLF at $z\sim6$ by combining our faint quasar sample with the bright quasar sample by SDSS and CFHQS.
Although our QLF estimate at $M_{1450}\sim-23$ is consistent with the previous study, including our data points invokes a higher number density at the faintest bin of the QLF than the previous estimate, making a steeper $\alpha$ and brighter $M_{1450}^*$.

5. The steepening of the QLF at the faint end increases the expected emission rate of ionizing photons; however, it only changes by a factor of approximately two to six, which was found to be still insufficient for the required photon budget of reionization.

\vskip 0.5cm
We could not conclude whether ELAIS109100446 is a quasar or a galaxy from the current observations.
Although it is challenging, follow-up NIR spectroscopy for other broad emission lines, such as C {\sc iv}, C {\sc iii} and Mg {\sc ii}, would draw more direct evidence of AGNs, distinguishing them from LBGs.  
More importantly, the line width of Mg {\sc ii} of our low-luminosity quasar sample will provide BH mass measurements toward a lower mass.
This will help us to understand the growing-up phase of SMBHs in the early universe.
Follow-up NIR spectroscopy will also determine chemical abundances in the broad-line regions of quasars.
Low-luminosity quasars sometimes show different features from bright quasars, such as prominent N {\sc iv}$]\lambda1486$ emission \citep{gli07} or broad N {\sc v} $\lambda1240$ emission \citep{mah05}.
It is surprising that the $z\sim7$ quasar spectrum has a strikingly good fit to that of low $z$ \citep{mor11, der13}, suggesting that this is not yet the first quasar.    
Our faint quasar sample would host SMBHs with lower masses, and its chemical abundance is interesting in the context of the initial star formation history and SMBH evolution.   
The submillimeter and radio follow-up observations will detect CO emission even from $z\sim6$ quasars \citep{wan13}, thereby revealing their dynamical masses and velocity structure.
Follow-up deep imaging will enable us to reveal galaxy clustering around the quasar, which has been used as a useful probe of possible sites of galaxy overdensity at high $z$, providing an intriguing clue to the nature of the first sites of co-evolution of quasars and galaxies and possible different star formation histories in high-density environments \citep{kas07}.

Finally, this observation is a good precursor study of the forthcoming Hyper Suprime-Cam (HSC) survey on the Subaru telescope.
The HSC survey consists of three-layer (wide: 1400deg$^2$, $r\lesssim26$; deep: 27deg$^2$, $r\lesssim27$; ultradeep: 3.5deg$^2$, $r\lesssim28$), multi-band imaging surveys in which $Y$-band, whose bandpass is close to the $z_R$ band in this study, imaging is planned.
The expected limiting magnitude on the wide layer is $Y_{\rm lim}=24.4$, which is slightly deeper than this study.
It also helps to effectively select quasars at $z\sim6$ as well as $\sim7$.
The expected number of quasars is $\sim350$ down to $M_{1450}=-23$ and $\sim50$ down to $M_{1450}=-24$ at $z\sim6$ and $7$, respectively.
We expect that HSC high-$z$ quasar survey will overcome current technical difficulties by combining a large telescope, a wide-field imaging camera ($\sim1$ deg$^2$), and an effective selection technique to systematically trace the QLF evolution beyond the knee up to $z\sim7$.  

\acknowledgments

We thank the referee for helpful comments that improved the manuscript.
We thank Toru Yamda for providing us the part of their $i$-band data. 
We thank Tohru Nagao, Hiroyuki Ikeda, Mattew Malkan, and Linhua Jiang for their useful discussions.
This work is based on data collected at the Subaru Telescope, which is operated by the National Astronomical Observatory of Japan.
We are grateful to the Subaru Observatory staff for their help with the observations.
We especially thank Miki Ishii and Takashi Hattori with help on Suprime-Cam and FOCAS observing and its data reduction.
This research was supported by the Japan Society for the Promotion of Science through Grant-in-Aid for Scientific Research 23340050.
This work is based in part on data obtained as part of the UKIRT Infrared Deep Sky Survey. 
This study is based on observations obtained with MegaPrime/MegaCam, a joint project of CFHT and CEA/DAPNIA, at the Canada–France–Hawaii Telescope (CFHT) which is operated by the National Research Council (NRC) of Canada, the Institut National des Sciences de l'Univers of the Centre National de la Recherche Scientifique (CNRS) of France, and the University of Hawaii. This work is based in part on data products produced at TERAPIX and the Canadian Astronomy Data Centre as part of the Canada–France–Hawaii Telescope Legacy Survey, a collaborative project of NRC and CNRS.



{\it Facilities:} \facility{Subaru (Suprime-Cam, FOCAS)}.




\clearpage



\end{document}